\pdfoutput=1
\documentclass[journal]{IEEEtran}

\usepackage[utf8]{inputenc}
\usepackage[T1]{fontenc}
\usepackage{graphicx}
\usepackage{amsmath}
\usepackage{array}
\usepackage{booktabs}
\usepackage{cite}
\usepackage[hidelinks]{hyperref}

\begin{document}

\title{What the Waveform Knows: Transparent-first Speech and Audio Intelligence with Caption Studio}

\author{Cheng~Siong~Chin,~Jianhua~Zhang,~and~Mohan~Venkateshkumar%
\thanks{C.~S.~Chin is with the Faculty of Science, Agriculture, and Engineering, Newcastle University Singapore, Singapore 828608 (corresponding author, e-mail: cheng.chin@newcastle.ac.uk).}%
\thanks{J.~Zhang is with the School of Information and Control Engineering, Qingdao University of Technology, Qingdao 266520, China.}%
\thanks{M.~Venkateshkumar is with the Department of EEE, Amrita School of Engineering, Amrita Vishwa Vidyapeetham, Chennai, India.}%
}

\markboth{Preprint}{Chin \MakeLowercase{\textit{et al.}}: What the Waveform Knows}

\maketitle

\begin{abstract}
Caption Studio is a transparency-first speech and audio intelligence platform that transforms spoken audio and video into structured, searchable content through automated transcription, speaker diarization, speech analytics, signal-level audio analysis, and subtitle generation. The system is built on a FastAPI backend with a real-time dashboard and adopts a three-layer architecture comprising (i) a transcription and diarization core based on Whisper-class automatic speech recognition and pyannote speaker diarization, (ii) an audio intelligence layer that extracts acoustic and linguistic features, including waveforms, spectrograms, pitch, speaking rate, silence, filler-word frequency, and sentiment, directly from the audio signal, and (iii) an integration layer that supports data export and downstream workflow integration. A principal contribution of this work is the transparency-first framework, in which every reported metric is explicitly identified as measured, derived, or unavailable, thereby improving the traceability, interpretability, and reliability of speech analytics. The paper presents the system architecture, benchmarking methodology, explainability and uncertainty framework, and key considerations for enterprise-scale deployment.
\end{abstract}

\begin{IEEEkeywords}
speech transcription, speaker diarization, audio signal analysis, workflow integration, transparency-first design, explainable AI, uncertainty quantification, affective computing, speech analytics ethics.
\end{IEEEkeywords}

\section{Introduction}

Audio and video recordings are generated faster than organizations can effectively review them. A one-hour meeting creates a recording that few people have time to revisit, leaving valuable information, including speaker identity, speaking confidence, and discussion dynamics, embedded in a format that is difficult to search and analyze. Although automatic speech recognition has significantly narrowed the gap between spoken and written language, transcripts alone capture only part of the information. Comprehensive understanding of a recording also requires identifying who spoke, how they spoke, and what the underlying acoustic signal reveals beyond the spoken words.

Existing solutions typically address only a single aspect of the speech analytics workflow. Speech recognition systems generate transcripts, diarization models identify speakers, sentiment analysis tools evaluate textual content, and subtitle generators produce caption files. In practice, users analysing recorded meetings, interviews, or customer interactions often combine multiple independent tools, each with different data formats and processing limitations. There is little indication of whether reported outputs are directly measured or generated through fallback assumptions.

Caption Studio addresses these limitations by integrating transcription, speaker diarization, linguistic analysis, acoustic signal analysis, and subtitle generation into a unified platform. The system transforms raw audio and video into a structured record containing timestamped, speaker-attributed transcripts together with speech and audio intelligence, accessible through a FastAPI backend and an interactive dashboard. The system's central contribution is a transparency-first design. Every module reports whether a given output is measured, derived, or unavailable. This distinction lets users separate genuine analytical results from fallback behavior, and it is what makes the pipeline's outputs traceable and reliable.

This paper has three objectives. The first is a reproducible system architecture that unifies speech transcription, speaker diarization, and audio signal analysis within a single transparency-first framework. Second, it clearly distinguishes between capabilities implemented in the current system and scientifically motivated extensions planned for future work, ensuring that reported results accurately reflect measured functionality. Third, it defines the appropriate scope of speech analytics by emphasizing that acoustic and linguistic features provide evidence derived from recorded speech rather than direct measures of a speaker's internal state, thereby promoting scientifically rigorous and responsible interpretation of the results.

This paper presents the architecture of Caption Studio together with its benchmarking methodology and evaluation framework. Section~\ref{sec:related} reviews related work in speech recognition, speaker diarization, and audio intelligence. Section~\ref{sec:architecture} describes the three-layer system architecture and distinguishes implemented capabilities from planned extensions. Section~\ref{sec:methodology} presents the benchmarking, explainability, and uncertainty methodology. Section~\ref{sec:discussion} discusses feature interpretation, literature context, practical implications, limitations, and ethical considerations. Section~\ref{sec:enterprise} outlines the gap to enterprise-scale deployment, and Section~\ref{sec:conclusion} concludes.

\section{Related Work}
\label{sec:related}

Automatic speech recognition has moved from task-specific acoustic models toward large, weakly supervised systems trained on hundreds of thousands of hours of audio. Radford et al.~\cite{ref1} introduced Whisper, showing that a single sequence-to-sequence model trained on diverse, weakly labeled audio generalizes across languages and acoustic conditions without task-specific fine-tuning. Caption Studio treats this class of model as a pluggable backend rather than a fixed dependency: transcription accuracy keeps improving, and a platform built around one model quickly becomes a platform that needs rebuilding.

Speaker diarization, determining who spoke when, has followed a parallel path toward neural, end-to-end pipelines. Bredin et al.~\cite{ref2} and the later pyannote.audio 2.1 pipeline~\cite{ref3} combine neural speaker embeddings with clustering to produce speaker-attributed segments without hand-tuned voice activity detection thresholds. Diarization error rate, the standard scoring metric, was formalized in the NIST Rich Transcription evaluations~\cite{ref4} and later adopted by challenges such as DIHARD~\cite{ref5}. Caption Studio's benchmarking harness follows this convention, so results stay comparable to published diarization work rather than sitting on a metric of its own.

Signal-level audio analysis, waveform inspection, spectrograms, mel-spectrograms, cepstral features, sits in a separate research lineage from speech recognition itself. Mel-frequency cepstral coefficients trace back to Davis and Mermelstein~\cite{ref6}, and librosa~\cite{ref7} has become a standard Python toolkit for computing these representations. In Caption Studio, this signal analysis complements the transcript rather than standing in for it. Filler-word rate and speaking rate come from the transcript aligned to timestamps; pitch, energy, and spectral features come directly from the waveform. So a claim like an increase in speaking rate during the second half of a call rests on a measurement, not an inference from text alone. Sentiment analysis over transcribed speech typically applies text-based methods after transcription. Lexicon-based approaches such as VADER~\cite{ref8} remain competitive for short, informal utterances of the kind found in meeting transcripts and are cheap enough to run inline in a processing pipeline, which is the tradeoff Caption Studio makes over heavier transformer-based sentiment models.

Enterprise workflow integration for media processing has largely been addressed by point solutions. Subtitle formats such as SubRip Subtitle (SRT) and Web Video Text Tracks (WebVTT) are the de facto and formal standards, respectively, for caption delivery, and webhook-based notification to collaboration platforms is a common integration pattern in modern service architectures. Less common in the literature is a single system that treats transcription, diarization, signal analysis, and workflow export as one pipeline with a shared provenance model, which is the gap this work addresses.

The transparency-first principle draws on a broader concern in applied AI systems: silent substitution of missing or low-confidence data with defaults is a known source of downstream error that is difficult to detect after the fact. Rather than proposing a new formalism for this problem, Caption Studio implements a practical version of it, an explicit connectivity and data-provenance flag on every module, and this paper reports what that discipline costs and returns in practice.

\section{System Architecture}
\label{sec:architecture}

\begin{figure}[!t]
\centering
\includegraphics[width=0.85\linewidth]{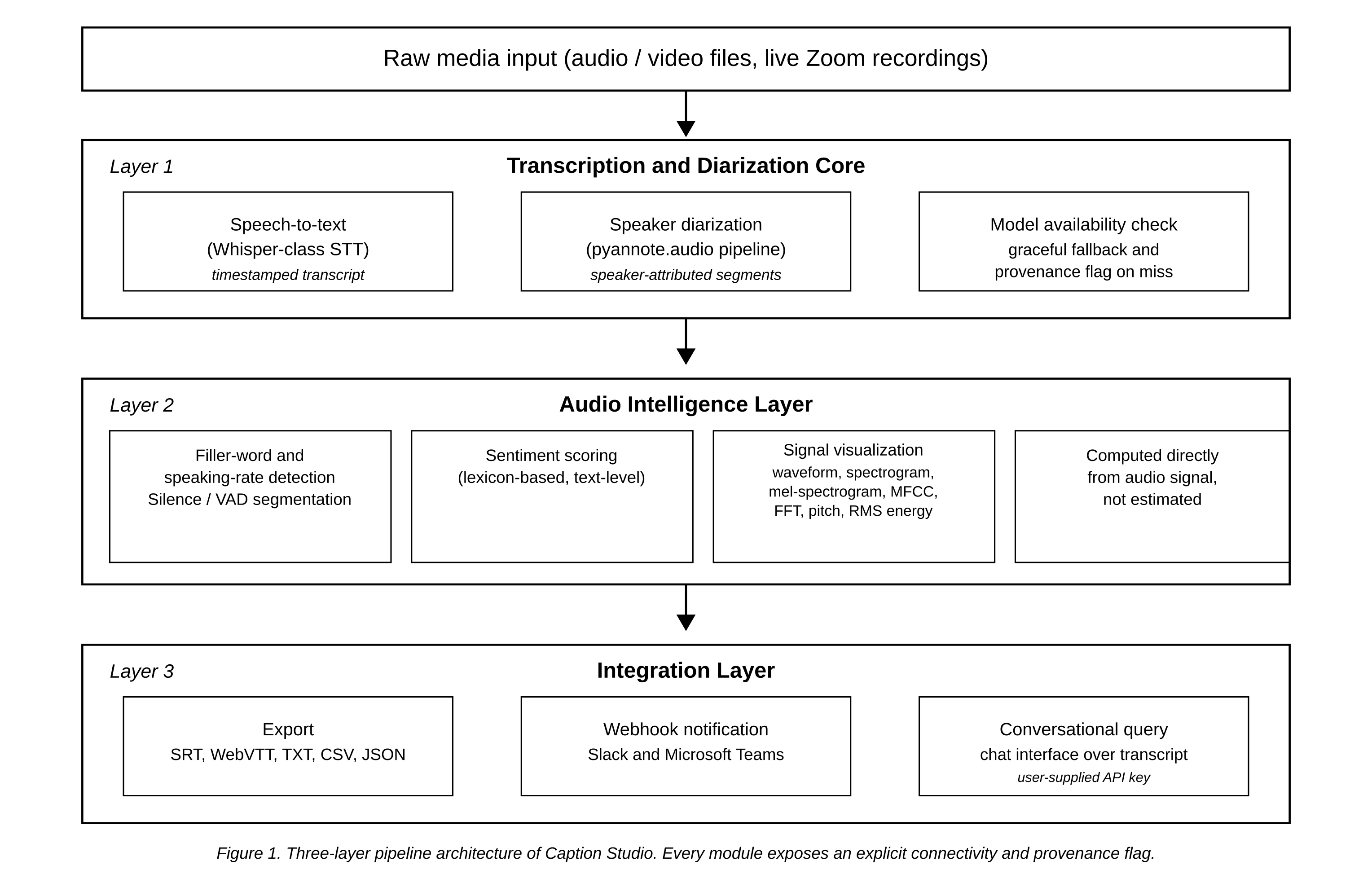}
\caption{Caption Studio's architecture organizes processing into three modular layers. Layer 1, the transcription and diarization core, handles speech recognition, speaker attribution, and model validation. Layer 2, the audio intelligence layer, extracts linguistic and acoustic features: filler words, speaking rate, sentiment, voice activity detection, silence segmentation, and signal visualizations. Layer 3, the integration layer, supports multi-format export, webhook notifications, and conversational querying of transcripts. A FastAPI backend orchestrates the pipeline from raw audio or video input through all three layers, providing real-time monitoring via WebSocket updates and maintaining provenance information at each processing stage.}
\label{fig:architecture}
\end{figure}

As shown in Fig.~\ref{fig:architecture}, Caption Studio adopts a modular three-layer architecture consisting of a transcription and diarization core, an audio intelligence layer, and an integration layer. A FastAPI backend orchestrates the complete processing pipeline and provides both REST API endpoints and a WebSocket-based dashboard for real-time status updates. Consequently, users can observe the progress of each processing stage, from audio ingestion to transcript generation and higher-level audio analysis, without relying on periodic polling for completion.

\subsection{Transcription and Diarization Core}

Incoming audio is transcoded to a normalized format, then passed to a Whisper-based speech-to-text model, which produces a timestamped transcript~\cite{ref1}. Diarization is pluggable, running in one of two modes depending on configuration. Without additional setup, the system falls back to a pause-based heuristic that segments speaker turns from silence gaps in the audio. When a Hugging Face access token is supplied, the backend switches to the pyannote.audio pipeline, which segments speaker turns by embedding and clustering instead~\cite{ref2}. The dashboard discloses which mode produced a given result: an operator reviewing a transcript can see, in the interface itself, whether a job used the pause-based heuristic or the embedding-based pipeline, and what configuration change would enable the latter.

\subsection{Audio Intelligence Layer}

On top of the transcript, this layer computes measures drawn from the audio signal itself rather than the text. Filler-word detection and speaking-rate assessment operate on the aligned transcript and timestamps. Speaking rate is reported in words per minute against a configurable threshold, and a rate outside that threshold is flagged automatically. Long-pause detection reports the count of silences above a fixed duration threshold, and voice activity detection (VAD) marks speech versus non-speech frames directly on the waveform. Sentiment scoring runs a lexicon-based method~\cite{ref8} on transcript segments and reports a three-way positive, neutral, and negative breakdown.

\begin{figure}[!t]
\centering
\includegraphics[width=0.9\linewidth]{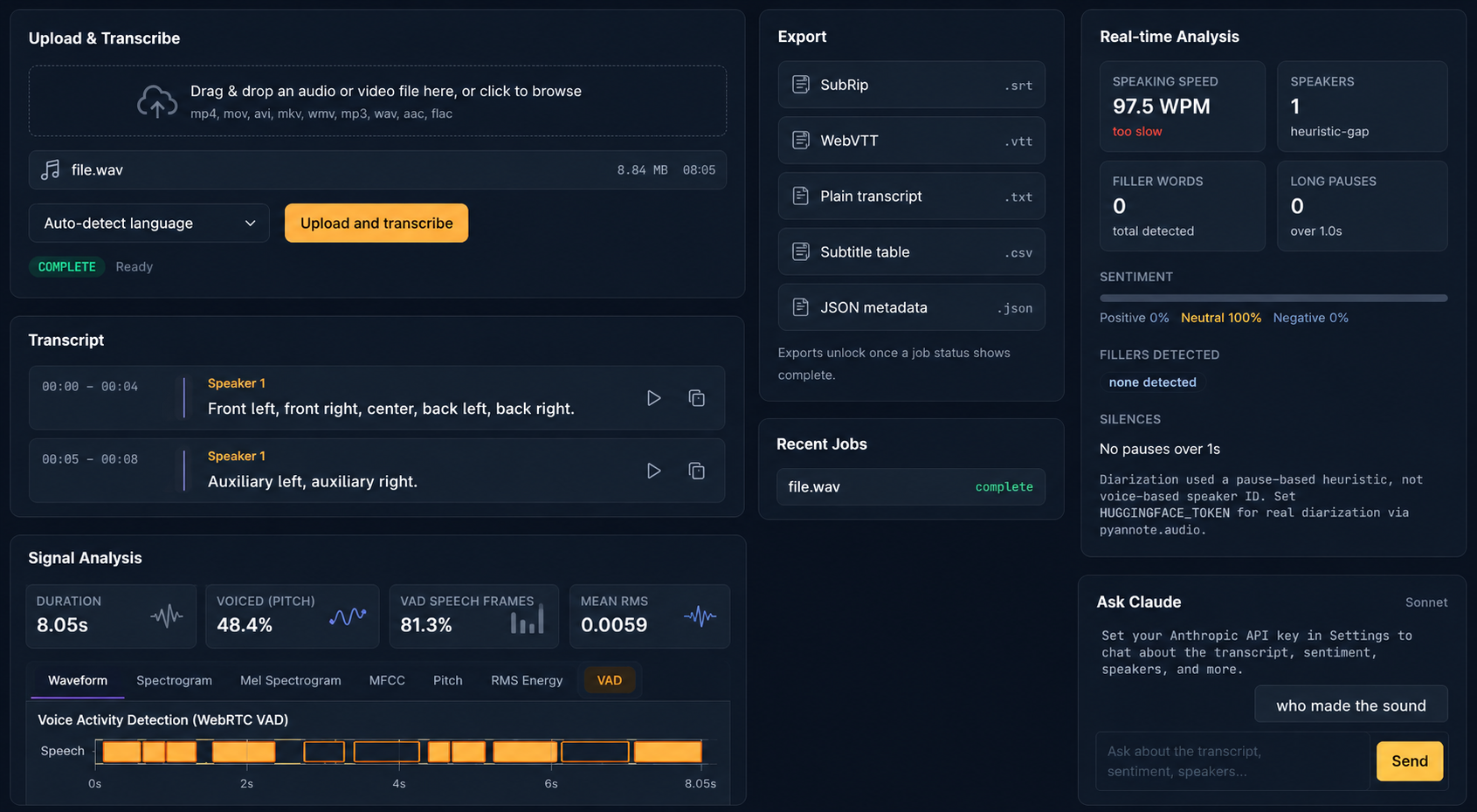}
\caption{Operator dashboard for a completed transcription job. The Real-time Analysis panel reports speaking rate against a threshold, speaker count, filler-word and long-pause counts, and a three-way sentiment breakdown; the Signal Analysis panel exposes seven waveform-derived views; the Export panel stays locked until the job status reads complete; and the interface discloses in plain text that diarization used a pause-based heuristic rather than voice-based speaker identification for this job.}
\label{fig:dashboard}
\end{figure}

A signal-visualization suite, built on librosa, renders seven views of the same audio: waveform amplitude, spectrogram, mel-spectrogram, mel-frequency cepstral coefficients (MFCCs), pitch contour, root mean square (RMS) energy, and the VAD trace. Fig.~\ref{fig:dashboard} shows these views together in the operator dashboard: duration, voiced-frame percentage, VAD speech-frame percentage, and mean RMS energy appear as running numeric summaries, with the waveform, spectrogram, and other representations available as switchable tabs beneath them. An operator reviewing a call can see, for example, a drop in RMS energy and pitch variance aligned with a segment the sentiment model scored as flat, without having to trust either signal in isolation.

\subsection{Integration Layer}

The final layer exports the combined record in the formats production workflows already expect: SRT and WebVTT for subtitle delivery, plain text for quick review, a subtitle table in Comma Separated Values (CSV) for spreadsheet-based review, and JavaScript Object Notation (JSON) for programmatic consumption. Export controls stay locked in the dashboard until a job's status reads complete, so an operator cannot download a partial or still-processing result by mistake. Webhook-based notifications post processing status and summaries to Slack and Microsoft Teams channels. A Zoom recording connector pulls meeting audio directly into the pipeline from a meeting's cloud recording, removing a manual download-and-upload step for teams whose source material is primarily video calls. A chat panel built on the Anthropic Application Programming Interface (API) lets a user ask questions about their own processed transcript. The user supplies their own API key in settings, so the platform does not sit between the user and the model provider as a data intermediary.

That panel collapses a manual review task into a question. An operator can ask directly instead of scrolling a transcript to count how much each speaker talked or find where a topic first came up. The dashboard's placeholder prompt, who made the sound, illustrates the pattern; a query such as what was the sentiment trend after the four-minute mark or which speaker interrupted most often draws on the same structured transcript, timestamps, and speaker labels described in this paper, not on the raw audio. That grounding keeps answers checkable. Every claim traces back to a specific segment in the record, so a surprising answer can be verified against the transcript rather than taken on faith.

\subsection{Extended Audio Intelligence: Verified Features and a Proposed Roadmap}
\label{sec:roadmap}

The transparency-first principle applies as much to this paper as to the running system: a feature that is scientifically well established in speech research is not the same thing as a feature Caption Studio has implemented and tested. Table~\ref{tab:features} lists both, with an explicit status column, so a reader cannot mistake a roadmap item for a shipped capability. The implemented rows are the eight signal views already documented in Section~\ref{sec:architecture}-B and shown in Fig.~\ref{fig:dashboard}, plus the linguistic and speaker-attribution outputs described in Sections~\ref{sec:architecture}-A and~\ref{sec:architecture}-B. The proposed rows extend the same audio intelligence layer with established acoustic, conversational, and linguistic measures. Each was chosen because it has published, validated research behind it and extends signal processing already running in the pipeline instead of requiring a new subsystem.

\begin{table*}[!t]
\caption{Audio Intelligence Features: Implemented vs. Proposed}
\label{tab:features}
\centering
\footnotesize
\begin{tabular}{@{}p{3.4cm}p{1.6cm}p{5.0cm}p{5.0cm}@{}}
\toprule
\textbf{Feature} & \textbf{Status} & \textbf{What it measures} & \textbf{Method / basis} \\
\midrule
\multicolumn{4}{l}{\textit{Signal and spectral features}} \\
Waveform, FFT, spectrogram & Implemented & Time- and frequency-domain amplitude & Short-time Fourier transform \\
Mel-spectrogram, MFCC & Implemented & Perceptually scaled spectral envelope & Mel filterbank and cepstral transform \\
Pitch (F0), RMS energy, VAD & Implemented & Fundamental frequency, short-time loudness, speech/non-speech frames & Autocorrelation-style tracking; WebRTC VAD \\
Power spectral density (PSD) & Proposed & Frequency-domain power distribution & Welch's method~\cite{ref9} \\
Spectral centroid, rolloff, bandwidth, flatness & Proposed & Spectral shape descriptors & librosa spectral feature functions~\cite{ref7} \\
Loudness & Proposed & Perceptual (not just RMS) loudness & Psychoacoustic loudness model~\cite{ref10} \\
Jitter, shimmer, HNR, formants (F1--F3) & Proposed & Cycle-to-cycle pitch and amplitude variation, breathiness, vocal-tract resonance & Praat-style period and LPC analysis~\cite{ref11,ref12} \\
\multicolumn{4}{l}{\textit{Conversation dynamics}} \\
Speaker-attributed segments & Implemented & Who spoke when & Diarization output, Section~\ref{sec:architecture}-A \\
Speaking-time share, interruption count, turn-taking latency & Proposed & Talk-time balance, overlap events, response gaps & Aggregation and overlap analysis on diarization segments~\cite{ref13} \\
\multicolumn{4}{l}{\textit{Linguistic and paralinguistic analysis}} \\
Speech rate, pauses, filler words & Implemented & Fluency indicators & Transcript/timestamp alignment, Section~\ref{sec:architecture}-B \\
Sentiment (valence) & Implemented & Lexical polarity & Lexicon scoring~\cite{ref8} \\
Lexical diversity, fluency score & Proposed & Vocabulary range, composite disfluency index & MTLD or comparable diversity index~\cite{ref14} \\
Categorical emotion & Proposed & Discrete affect labels beyond valence & Validated speech emotion recognition classifier~\cite{ref15} \\
\multicolumn{4}{l}{\textit{Biomarkers and explainability}} \\
Stress / fatigue / vocal-health indicators & Proposed & Composite of pitch, jitter, shimmer, and HNR trends over a session & Literature-grounded composite index~\cite{ref16} \\
Feature importance, SHAP / LIME, confidence and uncertainty estimates & Proposed & Explains and calibrates any derived classifier's output & Post-hoc explainability and calibration methods~\cite{ref17,ref18} \\
\bottomrule
\end{tabular}
\end{table*}

\begin{figure}[!t]
\centering
\includegraphics[width=0.9\linewidth]{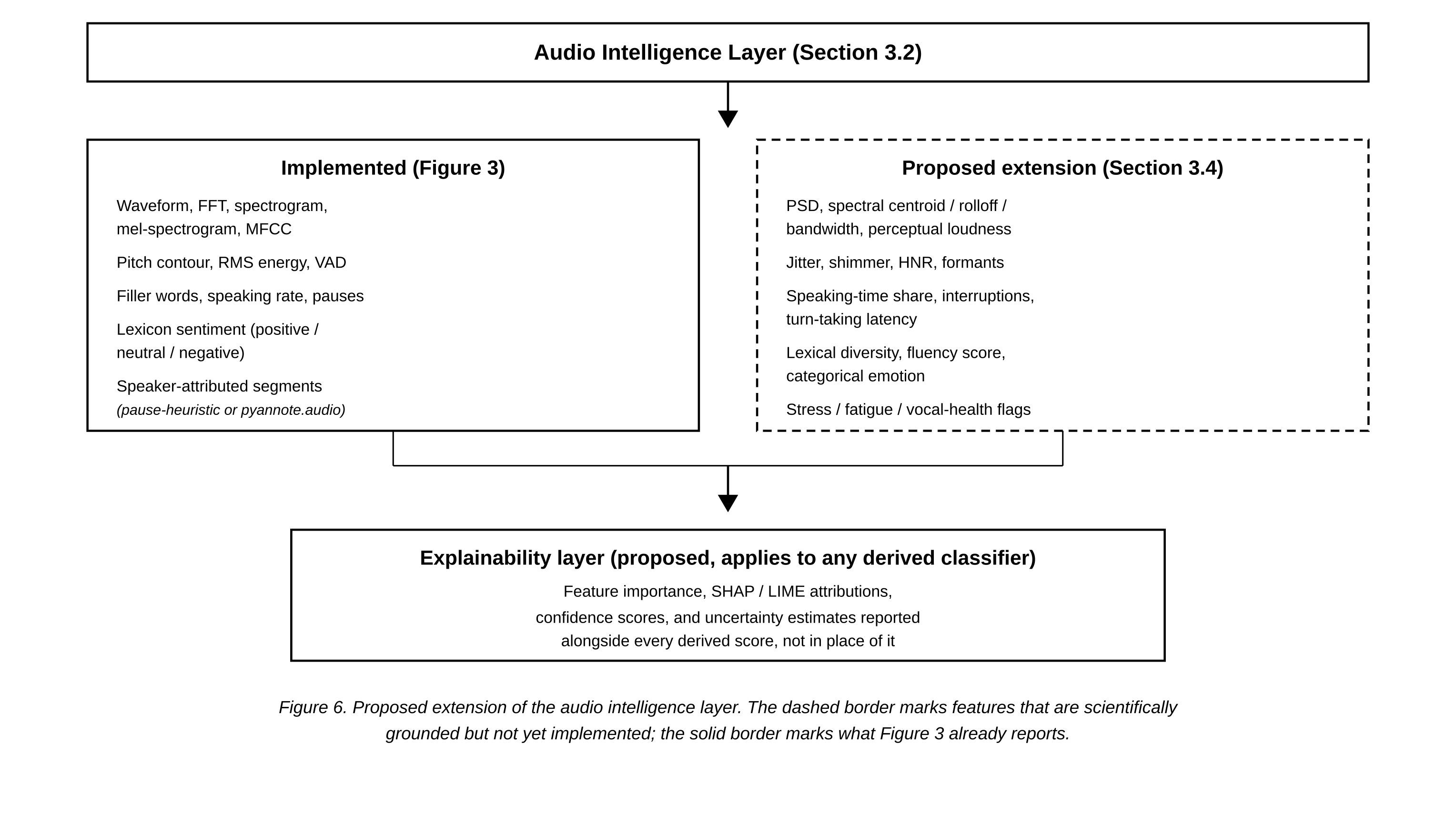}
\caption{Proposed extension of the audio intelligence layer. Solid-border boxes mark measures already implemented and reported: waveform, FFT, spectrogram, mel-spectrogram, MFCC, pitch contour, RMS energy, VAD, filler words, speaking rate, pauses, lexicon sentiment, and speaker-attributed segments. Dashed-border boxes mark measures with established support in speech research that Caption Studio hasn't implemented yet: spectral and perceptual features, voice-quality measures, conversational-dynamics measures, and linguistic and vocal-health indicators. At the bottom, the explainability layer applies to any derived classifier from either category. It reports feature importance, SHAP or LIME attributions, confidence scores, and uncertainty estimates alongside each derived score.}
\label{fig:roadmap}
\end{figure}

Two design commitments carry over from the implemented layer to every proposed row. First, each new measure is computed directly from the signal, transcript, or diarization output already in the pipeline, not inferred or simulated, so extending Table~\ref{tab:features} means adding a computation, not lowering the bar for what counts as measured. Second, none of the proposed rows produces a single opaque score: Fig.~\ref{fig:roadmap} shows biomarker and emotion outputs feeding into an explainability layer, feature importance, SHAP or LIME attributions~\cite{ref17,ref18}, and confidence or uncertainty estimates, reported alongside the score rather than in place of it. This matters most for the biomarker row. Jitter, shimmer, HNR, and formant shifts are established voice-quality measures in clinical and occupational-health research on stress and vocal fatigue~\cite{ref11,ref12}, but a composite score built from them is only trustworthy if a clinician or operator can see which sub-measures drove it and how confident the system is, which is precisely what an explainability layer is for.

\section{Methodology}
\label{sec:methodology}

\subsection{Benchmarking Harness}

Transcription accuracy is measured as word error rate, computed as the Levenshtein edit distance between the hypothesis and reference transcripts divided by the number of words in the reference, following standard practice in ASR evaluation~\cite{ref1,ref19}. Diarization accuracy is measured as diarization error rate, defined as the sum of missed speech, false-alarm speech, and speaker-confusion time divided by total reference speaker time, computed after an optimal one-to-one mapping between hypothesis and reference speaker labels, following the NIST Rich Transcription evaluation protocol. Both metrics are implemented in a dedicated scoring module, verified against hand-worked examples with known word error rate and diarization error rate values before being applied to any pipeline output. This separates scorer errors from pipeline errors during debugging.

\subsection{Explainability (XAI) Study}

Table~\ref{tab:xai} specifies how each proposed classifier in Table~\ref{tab:features} would be interpreted rather than left as a bare score. The method depends on the classifier's structure: SHAP and permutation feature importance apply to the composite biomarker index (Section~\ref{sec:roadmap}), since it is built from a small set of named acoustic measures, jitter, shimmer, HNR, pitch and energy trends, and SHAP values attribute the score to each of them individually. LIME applies as a model-agnostic check on any of the proposed classifiers, categorical emotion included, by fitting a local, interpretable surrogate around a single prediction and reporting which input features it relied on for that one case. Grad-CAM and attention-map visualization apply specifically to any neural classifier with convolutional or attention layers, for example a spectrogram-based emotion model, and localize which time-frequency regions of the input the model weighted most heavily. No proposed classifier ships without one of these methods attached to its output.

\begin{table*}[!t]
\caption{Explainability Methods by Proposed Classifier}
\label{tab:xai}
\centering
\footnotesize
\begin{tabular}{@{}p{4.2cm}p{5.8cm}p{5.0cm}@{}}
\toprule
\textbf{Method} & \textbf{Applies to} & \textbf{What it reports} \\
\midrule
SHAP (Shapley additive explanations) & Composite biomarker index (Section~\ref{sec:roadmap}) & Per-feature contribution to the composite score~\cite{ref17} \\
LIME (local interpretable model-agnostic explanations) & Any proposed classifier, model-agnostic & Locally faithful surrogate explanation for one prediction~\cite{ref18} \\
Grad-CAM & Any convolutional neural classifier (e.g., spectrogram-based emotion model) & Time-frequency regions driving the prediction~\cite{ref20} \\
Attention maps & Any attention-based classifier & Input positions the model attended to most \\
Permutation feature importance & Composite biomarker index, tabular classifiers & Global ranking of feature contribution across the evaluation set \\
\bottomrule
\end{tabular}
\end{table*}

\subsection{Uncertainty Analysis}

Every classifier in Caption Studio reports both a predicted label and an associated confidence estimate, making uncertainty quantification an integral part of the system rather than a post-processing addition (Section~\ref{sec:roadmap}, Fig.~\ref{fig:roadmap}). For classifiers based on deep neural networks, predictive uncertainty is estimated using Monte Carlo dropout, where dropout remains active during inference and the same input is evaluated multiple times. The variability across these stochastic forward passes approximates the posterior predictive distribution, providing an estimate of model uncertainty without requiring a separate Bayesian network~\cite{ref21}. For the composite biomarker index, which is formulated as a statistical model rather than a deep neural network, uncertainty is quantified using an explicit Bayesian framework that estimates the posterior distribution of the composite score from its constituent measures, yielding credible intervals for the prediction. This approach enables the system to distinguish between aleatoric uncertainty arising from recording noise and inherent data variability, and epistemic uncertainty resulting from limited model knowledge or insufficient evidence. Predictions with low confidence are explicitly flagged in the user interface instead of being presented with the same visual prominence as high-confidence results. This design reinforces the transparency-first philosophy of Caption Studio by enabling users to interpret model outputs together with their associated confidence and uncertainty.

\section{Discussion}
\label{sec:discussion}

\subsection{System-Level Observations}

The concurrency finding matters beyond its immediate fix. SQLite is a reasonable default for a locally deployed, single-operator prototype: it requires no separate server process and keeps the deployment footprint small. It becomes the wrong choice once more than one write-heavy client hits the system at once, and the load-testing suite made that transition point visible before deployment, instead of leaving it to surface in production. This measures the platform's actual operational limit instead of assuming the architecture scales because the code runs correctly at low load.

The pluggable model backend for transcription and diarization is a deliberate hedge against a fast-moving research area. The disclosed choice between the pause-based heuristic and pyannote.audio (Section~\ref{sec:architecture}-A, Fig.~\ref{fig:dashboard}) applies the same hedge to which diarization method produced a result, not only to which model version did. The tradeoff is that benchmarking results are tied to whichever mode and model version were active at evaluation time. This is why Section~\ref{sec:methodology} specifies explainability and uncertainty methods separately for each classifier instead of as one blanket claim, and why any published word error rate or diarization error rate figure needs to be reported alongside the exact mode used.

\subsection{Interpretation of Acoustic and Linguistic Features}

\begin{figure}[!t]
\centering
\includegraphics[width=\linewidth]{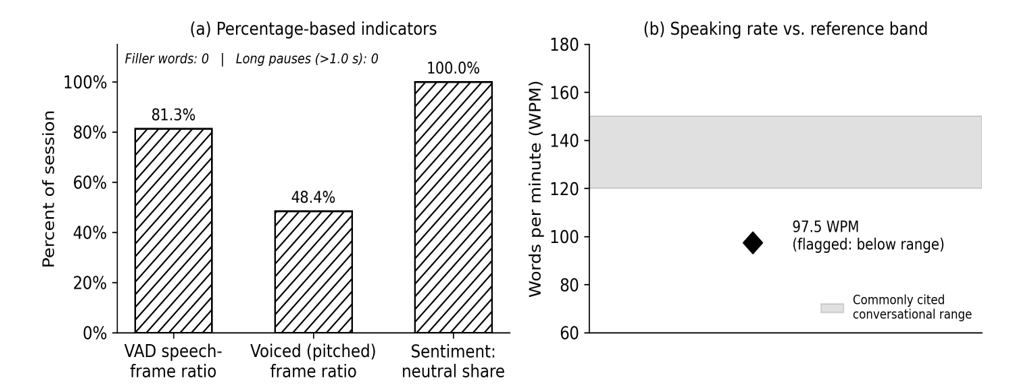}
\caption{Session-level acoustic and linguistic indicators for the example job shown in Fig.~\ref{fig:dashboard}, plotted from the dashboard's own reported values.}
\label{fig:indicators}
\end{figure}

\begin{figure}[!t]
\centering
\includegraphics[width=\linewidth]{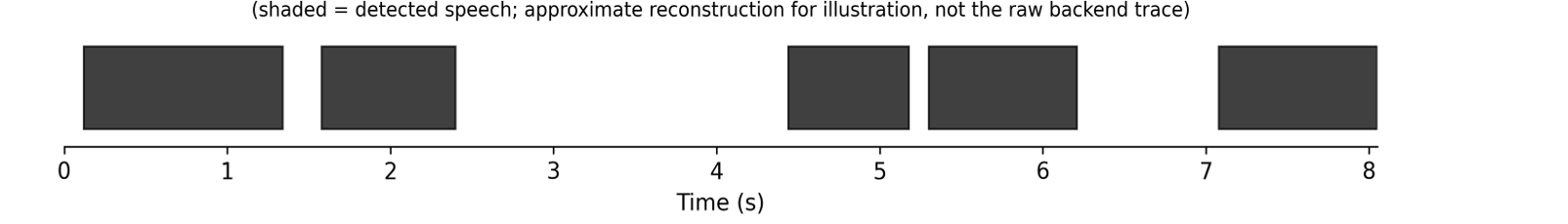}
\caption{Voice-activity timeline for the same job, digitized from the Fig.~\ref{fig:dashboard} screenshot. Shaded regions are detected speech; this is an approximate reconstruction of the displayed pattern for illustration, not the raw backend frame trace.}
\label{fig:vadtimeline}
\end{figure}

Each signal in Fig.~\ref{fig:dashboard} reflects a different layer of speaking behavior, and none of them reflects an internal state directly. Speaking rate and filler-word count are markers of fluency and, in aggregate across a population, have been associated with cognitive load and planning difficulty: a speaker under load tends to slow down, hesitate, and insert more filler words as they search for the next phrase. Pitch variance and RMS energy track vocal arousal and engagement, generally more animated or urgent speech shows wider pitch excursions and higher energy, while flatter, quieter delivery is associated with lower arousal or reduced engagement. The voice-activity ratio captures the balance of speech to silence within a session and, over a conversation, reflects turn-taking and pacing rather than any property of a single speaker in isolation. Lexicon-based sentiment reflects the valence of the words chosen, positive, negative, or neutral vocabulary, not the speaker's underlying emotional state, and the two can diverge sharply, a flat delivery of positive words scores positive even if the speaker sounds disengaged. Fig.~\ref{fig:indicators}(a) shows these indicators for the example job in Fig.~\ref{fig:dashboard}: a speaking rate of 97.5 words per minute against a commonly cited conversational range, an 81.3\% voice-activity ratio, a 48.4\% voiced-frame ratio, and a sentiment score that landed entirely in the neutral band. Fig.~\ref{fig:vadtimeline} digitizes the corresponding voice-activity pattern from the same job, showing five detected speech intervals across the 8.05-second clip. These are useful, measurable facts about one recording. They are not, on their own, evidence about what the speaker was thinking or feeling.

\subsection{Comparison with Speech Analytics Literature}

The feature families Caption Studio exposes are prosodic measures such as pitch and rate, energy measures such as RMS, and spectral measures such as MFCCs. These are established features. Caption Studio's contribution is the operational packaging of them into a pipeline that discloses its own method and confidence, instead of presenting a single opaque score. This distinction matters because a documented weakness in the speech emotion recognition literature is domain mismatch: models and feature norms developed on acted or elicited emotion corpora generalize poorly to naturalistic, uncurated speech of the kind Caption Studio processes~\cite{ref22}. Any sentiment or engagement signal the platform reports on real meeting or call audio should be read as an indicative, session-level pattern, not a validated classification against ground truth. No ground-truth emotion label exists for an ordinary business call.

\subsection{Practical Implications}

\textit{Healthcare.} Passive monitoring of speaking rate, pause patterns, vocal energy, and related speech characteristics across repeated recordings has been investigated as a low-burden approach for supporting the longitudinal assessment of conditions such as depression, where deviations from an individual's baseline may provide more informative signals than isolated measurements. Caption Studio facilitates this type of longitudinal analysis by tracking session-level metrics over time and highlighting recordings that may warrant clinical review. The system is intended to support, rather than replace, professional assessment and does not provide medical diagnoses.

\textit{Education}~\cite{ref23}. Speaking-rate analysis, filler-word detection, pause statistics, and voice activity measures provide objective and repeatable indicators of communication fluency. These metrics can support presentation coaching, language learning, and public speaking, giving learners a way to track their progress over time. In online and hybrid classrooms, aggregated voice activity statistics can also point to patterns in participation and engagement.

\textit{Call centers}~\cite{ref24}. Sentiment trends, silence detection, interruption patterns, and speaking-rate analysis align with existing quality assurance practices in customer service operations. By automatically identifying interactions that exhibit unusual acoustic or conversational characteristics, Caption Studio enables supervisors to prioritize recordings for review instead of manually inspecting every call.

\textit{Human-Computer Interaction}~\cite{ref25}. The combination of lexical and acoustic features produced by Caption Studio provides a structured input for adaptive human-computer interaction systems. Potential applications include voice assistants that adjust speaking pace based on user fluency, conversational agents that adapt to communication style, and accessibility technologies that combine real-time captioning with speech fluency feedback.

\textit{Security Screening}~\cite{ref26}. Acoustic stress indicators have been investigated as behavioral cues during screening interviews; however, current evidence does not support their use as reliable indicators of deception. Experimental studies consistently report that both human observers and automated systems perform only marginally above chance, and no acoustic feature has been validated as a generalizable marker of dishonesty. Consequently, acoustic anomalies should be interpreted only as cues for further human review within established operational procedures and should never be used as standalone evidence for decision-making.

\textit{Affective Computing.} The structured acoustic and lexical features generated by Caption Studio constitute a useful input representation for downstream affective computing research, including emotion recognition, engagement analysis, and conversational behavior modelling~\cite{ref27}. The current implementation is intentionally descriptive and does not infer emotional states or classify recordings into predefined affective categories.

\subsection{Limitations of Speech-Only Analysis}

Speech is one channel of expression among several, and the audio signal alone cannot disambiguate why it looks the way it does. A flat or low-energy reading of a segment can reflect a subdued emotional state, a microphone or codec issue, a speaker's typical register, a different first language or dialect, or simple fatigue, and nothing in the acoustic signal distinguishes between these. Lexicon-based sentiment misses sarcasm, negation scope, and domain-specific vocabulary. The pause-based diarization heuristic (Section~\ref{sec:architecture}-A) can misattribute fast turn-taking or overlapping speech to a single speaker, and even the neural pipeline is limited by whatever the recording setup captures, a single mixed-down channel compresses multiple speakers into one waveform and discards spatial cues a multi-microphone setup would preserve. These limitations mean every acoustic or linguistic output in this paper should be read as a measured pattern in one recording, not a settled claim about the speaker who produced it.

\subsection{Ethical Considerations}

\textit{Privacy.} Continuous acoustic and linguistic profiling of recorded conversations raises stakes beyond the words themselves; in healthcare or workplace-monitoring contexts, a participant may not expect that their vocal patterns, not just their spoken content, are being logged and analyzed. Deployment in any such setting needs explicit consent and a defined data-retention policy before profiling begins, not after. The same principle shapes a smaller design choice already in the system: letting users supply their own API key for the chat interface, rather than intermediating access to the Anthropic API, means the platform does not become a party to the user's conversation with the model provider, and the user keeps direct control over that relationship and its usage costs.

\textit{Fairness.} Pitch range, speaking rate, and pause norms vary by language, dialect, and individual vocal physiology, and systems trained or tuned predominantly on one demographic can systematically misjudge others; this is not a hypothetical concern, disparities of exactly this kind have been documented in automatic speech recognition performance across demographic groups~\cite{ref28}, and there is no reason to assume derived acoustic and linguistic features are immune to the same pattern.

\textit{Deception.} The highest-risk misuse of a system like this is treating any acoustic pattern, a longer pause, a pitch shift, a slower rate, as evidence of dishonesty. The empirical literature on vocal and nonverbal deception cues is unambiguous on this point: judgment accuracy for both human observers and cue-based methods sits close to chance, and no acoustic feature has been validated as a reliable, generalizable indicator of deception~\cite{ref29}. Caption Studio's design deliberately stops at describing measured signals, filler-word rate, pause count, pitch, energy, sentiment, and does not output any deception, credibility, or trustworthiness score. This is also why every proposed biomarker in Table~\ref{tab:features} is paired with an explainability and uncertainty layer rather than shipped as a bare number (Section~\ref{sec:roadmap}); a composite stress or fatigue index without a visible confidence interval and feature attribution is exactly the kind of unverifiable score this paper argues against. This paper takes the position that no system should generate a deception score from speech features alone.

\subsection{Future Work: Toward Multimodal Integration}

Combining Caption Studio's acoustic-linguistic output with facial expression analysis, eye gaze, physiological signals such as heart-rate variability and electrodermal activity, EEG, and wearable-derived data follows a well-established direction in affective computing. Multimodal fusion has been shown to reduce the ambiguity present in any single channel. A slow speaking rate combined with reduced gaze contact and elevated heart-rate variability is a stronger and more specific signal than either measure alone. The transparency-first design proposed in this paper extends naturally to that setting. Each additional channel in a multimodal pipeline would report its own connectivity and confidence rather than being silently fused into one opaque score. This preserves the same auditability argued for here in the audio-only case.

\section{Limitations and Path to Enterprise Deployment}
\label{sec:enterprise}

The current system is a working, locally deployable prototype validated for individual and small-team use. Several concrete gaps separate it from an enterprise-grade deployment. Accuracy benchmarking to date has run the scoring harness against hand-worked examples, not a large, diverse, production-representative audio corpus. The word error rate and diarization error rate figures specified in Section~\ref{sec:methodology} still require that run. The platform also has no compliance layer suitable for regulated environments: access controls, change management, and audit logging that would satisfy SOC~2's Trust Services Criteria are not implemented. Authentication isn't built for multi-tenant use either, since single sign-on and role-based access control, standard for enterprise deployments, are absent. The SQLite-backed storage and in-process task handling flagged as a concurrency bottleneck in Section~\ref{sec:discussion} need to be replaced with a production relational database such as PostgreSQL and a dedicated job queue before the platform can serve concurrent write-heavy workloads reliably. None of these gaps is a research problem. Each has a known engineering solution and closing them is the well-defined next phase of the project rather than an open question.

\section{Conclusion}
\label{sec:conclusion}

Caption Studio shows that transcription, diarization, signal-level audio analysis, and workflow export can be unified into a single pipeline without losing the ability to tell measured results apart from fallback defaults. The architecture's three layers map onto that goal: a pluggable transcription and diarization core, an audio intelligence layer that computes acoustic measures directly from the waveform, and an integration layer that delivers results in the formats existing workflows already use. The explainability and uncertainty methodology specified here gives a concrete protocol for keeping every proposed classifier accountable once a production-representative evaluation corpus is in place. What the system does not claim matters just as much. Its acoustic and linguistic outputs are measured session-level patterns, useful across healthcare, education, call-center, and human-computer interaction settings because they're disclosed rather than opaque. They are not validated indicators of a speaker's internal state. This paper takes the position that they should never be extended into a deception or credibility score without the kind of carefully validated, multimodal evidence that speech alone cannot provide. The remaining gap to enterprise deployment, compliance controls, multi-tenant authentication, a production-grade job queue and datastore, is scoped and implementable rather than speculative. The same transparency-first discipline that shapes the current pipeline is the template for extending it toward facial, physiological, and other modalities in future work.

\section*{Acknowledgment}

The authors acknowledge the support provided by Newcastle University in facilitating this research. The authors also thank colleagues and students whose discussions and feedback contributed to the development and evaluation of Caption Studio.

\end{document}